\documentclass[prl,amssymb,twocolumn,showpacs,superscriptaddress,floatfix]{revtex4}

\usepackage{graphicx}
\usepackage{bm}

\begin{document}

\title{Signature of the Ground-State Topology in the Low-Temperature
Dynamics of Spin Glasses}

\author{F. Rom\'a}
\affiliation{Centro At{\'{o}}mico Bariloche, Consejo Nacional de
Investigaciones Cient\'{\i}ficas y T\'ecnicas, 8400 San Carlos de
Bariloche, R\'{\i}o Negro, Argentina}
\affiliation{Departamento de
F\'{\i}sica, Universidad Nacional de San Luis, 5700 San Luis,
Argentina}
\author{S. Bustingorry}
\affiliation{Centro At{\'{o}}mico Bariloche, Consejo Nacional de
Investigaciones Cient\'{\i}ficas y T\'ecnicas, 8400 San Carlos de
Bariloche, R\'{\i}o Negro, Argentina}
\author{P. M. Gleiser}
\affiliation{Centro At{\'{o}}mico Bariloche, Consejo Nacional de
Investigaciones Cient\'{\i}ficas y T\'ecnicas, 8400 San Carlos de
Bariloche, R\'{\i}o Negro, Argentina}

\begin{abstract}
We numerically address the issue of how the ground state topology
is reflected in the finite temperature dynamics of the $\pm J$
Edwards-Anderson spin glass model. In this system a careful study
of the ground state configurations allows to classify spins into
two sets: solidary and non-solidary spins. We show that these sets
quantitatively account for the dynamical heterogeneities found in
the mean flipping time distribution at finite low temperatures.
The results highlight the relevance of taking into account the
ground state topology in the analysis of the finite temperature
dynamics of spin glasses.

\end{abstract}


\date{\today}

\maketitle

Spin glass models are the paradigm of disordered systems with
slow dynamics \cite{Mezard,Young,Binder}. The main ingredients which
define these models are quenched disorder and an inherent
frustration in the interactions. These ingredients lead to a
non-trivial ground state topology \cite{Binder}, and
slow dynamics with spatial and dynamical heterogeneities
\cite{Ricci,Castillo02,Chamon02,Castillo03,Chamon04,Montanari03a,Montanari03b}.
Works analyzing the out-of-equilibrium properties have intuitively
suggested a relation between dynamical heterogeneities and the
ground state topology \cite{Ricci,Kisker96,Barrat99}. However, a
quantitative understanding of this precise relation still remains
an open question.

In particular, recent works \cite{Montanari03a,Montanari03b} have
analyzed the  dynamical
heterogeneities found in three different heterogeneous spin models.
By studying single spin dynamics different qualitative behaviors
were observed. On the other hand, other studies
\cite{Castillo02,Chamon02,Castillo03,Chamon04} have focused on spatially
coarse grained quantities and analyzed heterogeneities within a given coarse
grained length.

In this work we take into account a global property, dictated by the ground
state topology, in order to analyze dynamical heterogeneities.
We establish for the first time a quantitative relation
between the ground state topology and the finite temperature
dynamical properties of a spin glass model. We find that the
dynamical heterogeneities are well accounted for by two sets of
spins characterized by their role in the ground state.

We consider in particular the two-dimensional $\pm J$
Edwards-Anderson (EA) spin glass model. In
this model there exist clusters of spins which maintain their
relative orientation for all configurations of the ground state
manifold \cite{Toulouse,Vannimenus,Barahona}.  We extend this ground state information
to analyze the behavior of the system at finite temperatures. In
order to do this we divide the system into two sets of spins:
 solidary spins, i.e. the spins that form these clusters, and non-solidary spins.
The consequences of this division are two-fold.  On one hand it
gives us a quantitative tool to establish a relation between the
ground state topology and the finite temperature dynamical
properties. On the other, it gives an intuitive physical frame in
which to interpret the results.

We begin our analysis considering the spin auto-correlation
function, which clearly illustrates the different  qualitative
behaviors observed when the proposed division is taken into
account. The non-solidary spins decorrelate faster than solidary
spins, which suggests a relation with the separation in fast and
slow degrees of freedom. In order to address this point we analyze
the time-scale separation as observed in the mean persistence
time and mean flipping time probability distribution functions
\cite{Ricci,Jung05}. We show that the observed time-scale separation can
be quantitatively accounted for by the two sets of spins.

We consider the two-dimensional $\pm J$ EA model for spin glasses
\cite{Mezard,Young,Binder}, defined on a square lattice with periodic boundary
conditions. The Hamiltonian of the model is
\begin{equation}
H = \sum_{\langle i,j \rangle} J_{ij} \sigma_{i} \sigma_{j}
\end{equation}
where $\sigma_i = \pm 1$ is the spin variable and $\langle i,j
\rangle$ indicates a sum over nearest neighbors. The coupling
constants $J_{ij} = \pm J$ are random variables chosen from a
bimodal distribution. The time evolution of the model is governed
by a standard Glauber Monte Carlo process with sequential random
updates.

\begin{figure}
\includegraphics[width=7cm,clip=true]{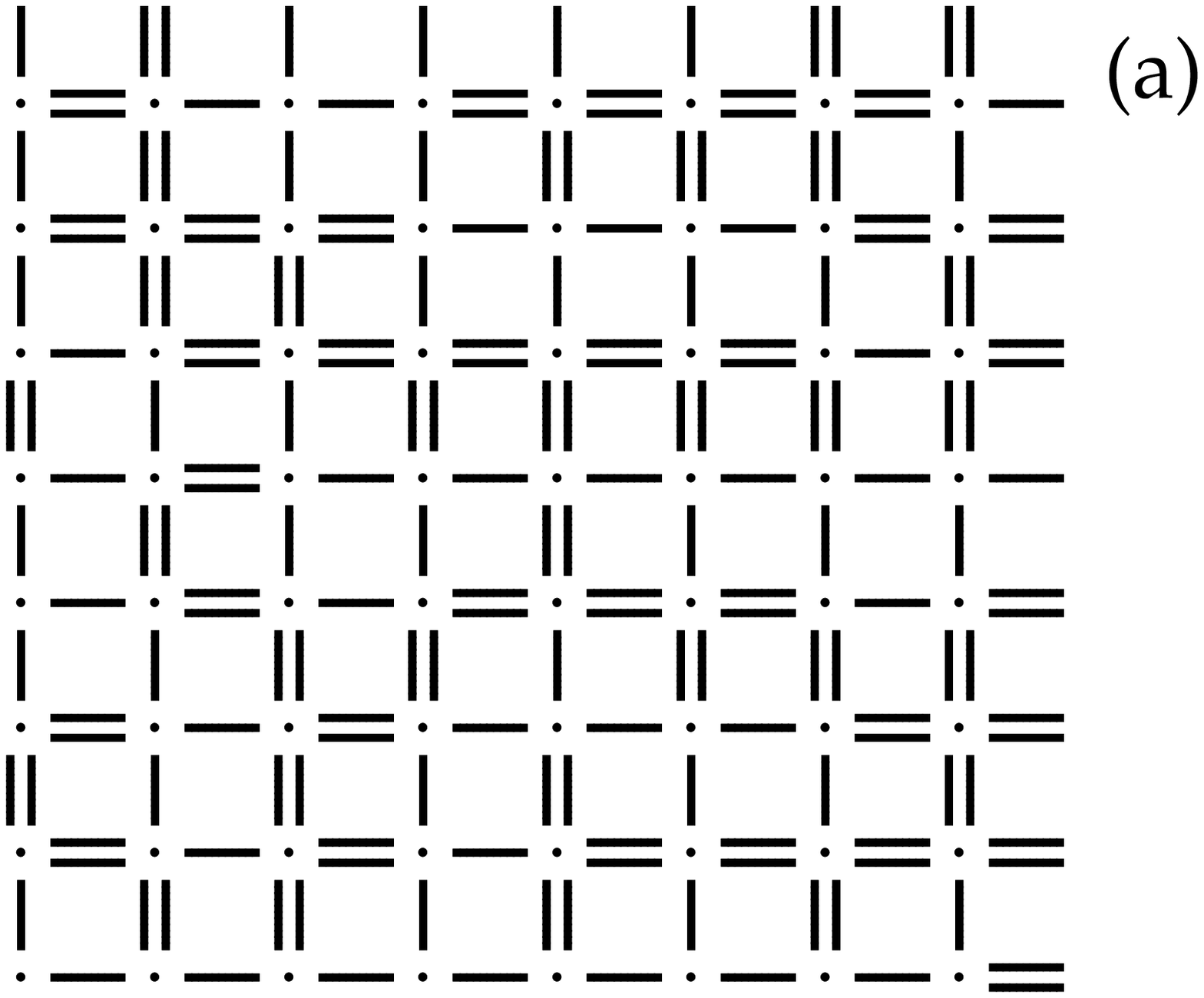}
\includegraphics[width=7cm,clip=true]{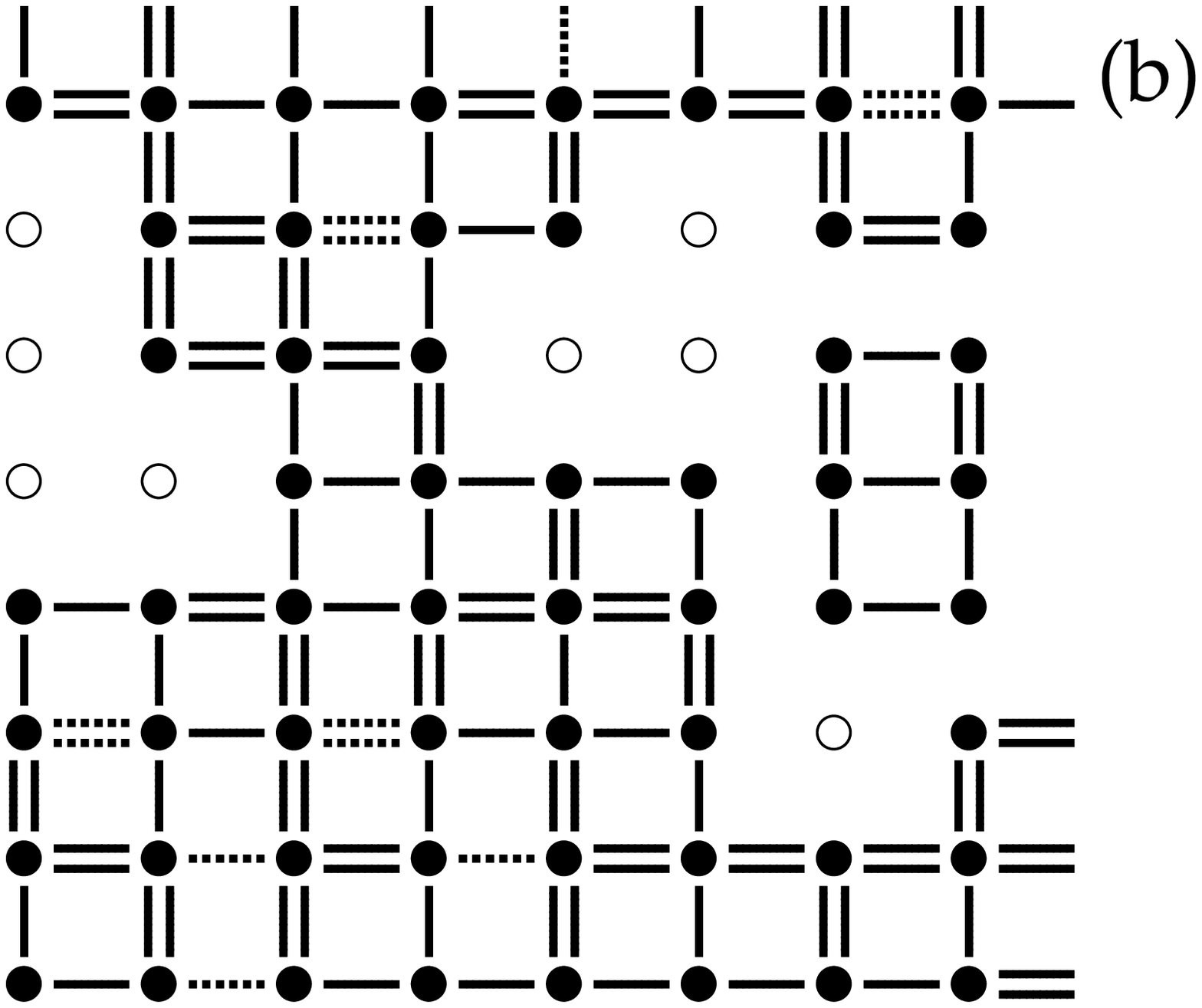}
\caption{\label{figure1} (a) A particular realization of bond
  disorder in an $8 \times 8$ lattice. Single (double) lines indicate
  ferromagnetic (anti-ferromagnetic) bonds.  (b) The corresponding rigid lattice (backbone).
  Full (dotted) lines indicate interactions which are always satisfied (frustrated) in
the ground state manifold. The solidary (non-solidary) spins are
indicated with close (open) circles.}
\end{figure}

In this model there exists clusters of {\em solidary} spins which maintain
their relative orientation for all configurations of the ground
state manifold \cite{Vannimenus,Barahona}. This backbone can be detected for each sample
through the identification of the diluted lattice \cite{Vogel98,Ramirez}, or
its generalization, the rigid lattice \cite{Barahona}.
The latter is formed by those bonds which are {\em always}
satisfied or {\em always} frustrated in the ground state manifold.
Notice that a backbone is also present in  other systems such as
the K-satisfiability model \cite{Barrat99,Monasson}.

A particular sample of size $N$ can be characterized by
recognizing all its solidary spins as shown in Fig. \ref{figure1}.
In order to obtain a statistical average over different
realizations of bond disorder, in all the results presented we
have calculated the sets of solidary spins for $2000$ different
samples in systems with size $16 \times 16$. All mean values are
obtained from averages with respect to both realizations of bond
disorder and thermal histories, as in Eq. \ref{corr}.

We begin our analysis of the out-of-equilibrium properties by
considering the two-time autocorrelation function,
\begin{equation}
C(t_w,t)=\frac{1}{N} \sum_{i=1}^{N} [ \langle \sigma_i(t_w)
  \sigma_i(t) \rangle ]
\label{corr}
\end{equation}
which measures the overlap of the spin configurations at times
$t_w$ and $t$ \cite{Cugliandolo}. The brackets $[\ldots]$ indicate
an average over different realizations of bond disorder, while
$\langle \ldots \rangle$ is a thermal average, i.e. an average
over different initial conditions and realizations of the thermal
noise. In each initial condition the spins take random values
$\sigma_i= \pm 1$, which  corresponds to a quench at $t=0$ from
$T=\infty$ to the temperature $T$ at which the system is analyzed.
It is worth stressing that usually one is interested in studying
the out-of-equilibrium properties below the critical temperature
$T_c$. However, in the two-dimensional EA spin glass model $T_c=0$
\cite{Roma,Rieger96}. Nevertheless it is widely accepted that for
low enough temperatures, the dynamics remains out of equilibrium
at short times and is very similar to the one observed in
three-dimensional spin glasses \cite{Barrat99,Barrat01}.

\begin{figure}
\includegraphics[width=7cm,clip=true]{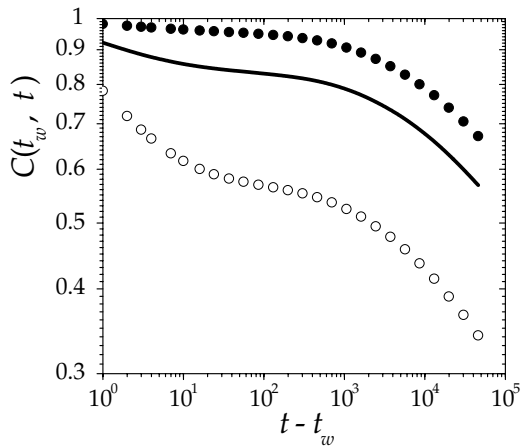}
\caption{\label{figure2} Autocorrelation function $C$ for
$t_w=10^4$ and $T=0.5$. The full line shows the behavior of  $C$
when all spins are taken into account. Close (open) circles
correspond to the behavior of the correlation $C_s(C_{ns})$ when
only solidary (non-solidary) spins are considered.}
\end{figure}

For each realization of bond disorder the division in solidary and
non-solidary spins can be taken into account rewriting the sum in
Eq. (\ref{corr}) as $C(t_w,t)=f_s C_{s} + f_{ns} C_{ns}$, where
$f_s$ ($f_{ns}=1-f_s$) is the fraction of solidary (non-solidary)
spins, and $C_{s}$ ($C_{ns}$) is the two-time autocorrelation
function restricted over the solidary (non-solidary) spins. Note
that the fraction of solidary spins is approximately $67 \% $ of
the total number of spins \cite{Roma}. Figure \ref{figure2} shows
the behavior of $C(t_w,t)$ vs. $t-t_w$ when $T=0.5$ and
$t_w=10^4$. For this parameters the system is in the aging regime
\cite{Barrat01}, and similar qualitative results are obtained for
lower temperatures. The full line corresponds to the behavior of
$C$ when all the spins are considered. The behavior of $C_{s}$
($C_{ns}$) is indicated with close (open) circles. For short times
($t-t_w < 10$) the solidary spins are strongly correlated, i.e.
they maintain their relative orientation in time. The non-solidary
spins present a qualitatively different behavior, with a  faster
decay of the correlation function. Only those spins which are
solidary in the ground state tend to remain correlated in time at
finite temperatures. For long times ($t-t_w
> 10^3$) each set of spins presents the same qualitative decay as
the whole system. This shows that the relaxation times of a
fraction of non-solidary spins is coupled to the ones of solidary
spins thus decorrelating together at longer times. This behavior
suggests a strong separation in characteristic times for the two
sets of spins and a possible path to analyze dynamical
heterogeneities as previously observed in Ref. \cite{Ricci}.

One possible path to the analysis of dynamical heterogeneities is
through the mean persistence time probability distribution
function (PDF). This quantity depends on the time window of interest,
given by $t_w$ and $t$, and is defined as the time at which, in
average, a given spin changes its state for the first time with
respect to its state at $t_w$. The mean persistence time, $\tau_p$,  is
obtained for every spin and the corresponding PDF is constructed,
$P_p(\ln \tau_p)$. The $\ln \tau_p$ scale is preferred due the
broadness of the PDF.

\begin{figure}
\includegraphics[width=7cm,clip=true]{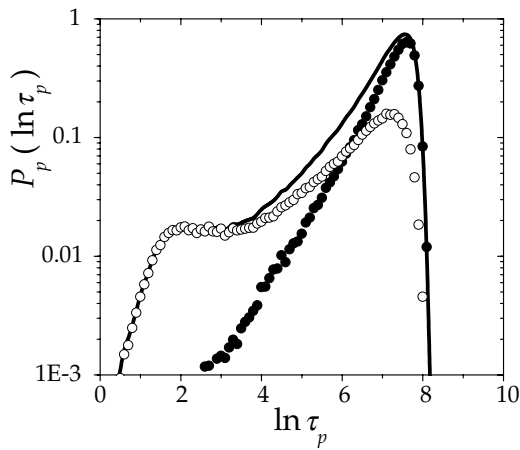}
\caption{\label{figure3} Mean persistence time PDF
for the time window $t-t_w=10^4$ with
$t_w=10^4$ and $T=0.5$. The whole distribution (full line) is divided
in the PDF of the solidary (close circles) and non-solidary (open
circles) spins.}
\end{figure}

In Fig. \ref{figure3} the behavior of $P_p(\ln \tau_p)$ is shown.
The symbols are the same as in Fig. \ref{figure2}. The PDF of the
whole system (full line) presents a sharp peak around $\ln \tau_p
\sim 7$ ($\tau_p \sim 10^3$), with a pronounced shoulder for lower
times. It is worth stressing that there exists a direct relation
between the mean persistence time PDF and the autocorrelation
function. For short times, the position of the shoulder, $\ln
\tau_p \sim 2$ ($\tau_p \sim 10$), corresponds to the first decay
observed in the full $C$, while for long times, the position of
the peak coincides with the second decay observed in the full $C$.
The division in solidary and non-solidary spins allows for a
physical interpretation of this time-scale separation. In Fig.
\ref{figure3} we show the mean persistence time PDF for solidary
and non-solidary spins separately. We observe that the shoulder
found in the full PDF is given only by a contribution of the
non-solidary spins. On the other hand, both solidary and
non-solidary spins contribute to the sharp peak. The
interpretation is straightforward: A fraction of non-solidary
spins decorrelate first due to their low mean persistence time. At
higher times the remaining fraction of the  non-solidary spins and
the solidary spins decorrelate together, both having similar mean
persistence time. The same relation between the mean persistence
time PDF and the autocorrelation  function was observed for lower
temperatures, giving support to our interpretation.

We expect this particular separation in solidary and non-solidary spins
to be reflected in other finite temperature dynamical quantities.
Recently, Ricci-Tersenghi and Zecchina \cite{Ricci} have observed a strong
time-scale separation in the mean flipping time PDF, $P_f$, as a signature of dynamical
heterogeneities. We analyze this quantity using the ground state information.
$P_f$ is obtained by measuring the number of flips ($N_{flips}$) done by every spin
within the time window extending from $t_w$ to $t$. The mean
flipping time $\tau_f$ for a given $t_w$ and $t$ is defined as the
time window size divided by the number of flips: $\tau_f= (t-t_w)/
N_{flips}$ \cite{Ricci}.

\begin{figure}
\includegraphics[width=7cm,clip=true]{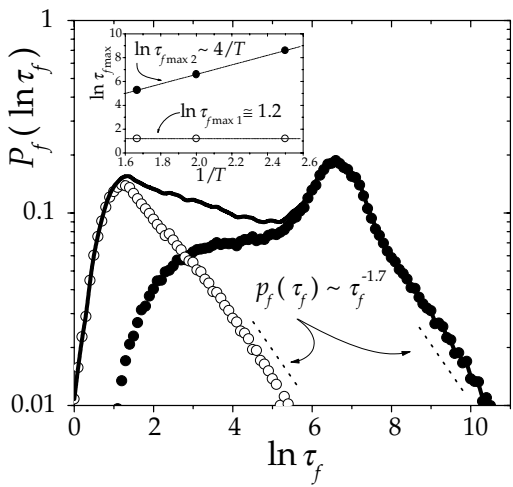}
\caption{\label{figure4} Mean flipping time PDF for the time
window $t-t_w=10^4$ with $t_w=10^4$ and $T=0.5$. The whole
distribution (full line) is divided in the PDF of the solidary
(close circles) and non-solidary (open circles) spins. The
power-law like behavior of the distribution's tails is
highlighted. The inset shows the behavior of the maximum of the
solidary (nonsolidary) spins PDF, $\ln\tau_{f\,\,\text{max2}}$
($\ln\tau_{f\,\,\text{max1}}$), for three different temperatures:
$T= 0.4$, $0.5$ and $0.6$.}
\end{figure}

Figure \ref{figure4} shows the behavior of $P_f(\ln \tau_f)$. The
symbols are the same as in  Fig. \ref{figure2}. The PDF of the
whole system (full line) presents two main peaks \cite{Ricci},
which are a manifestation of strong dynamical heterogeneities
\cite{Nota1}. Generally speaking these two peaks correspond to
fast (left peak) and slow (right peak) spins. We also measure the
mean flipping time distribution for solidary and non-solidary
spins separately. In Fig. \ref{figure4} we show that the two peaks
of the full PDF can be well accounted for by this separation. The
slow (fast) spins at finite temperature correspond to solidary
(non-solidary) spins. At high temperatures, the two peaks collide
and the strong time scale separation is no longer observable.

It is worth stressing that this separation reveals that a further
internal structure is present, as can be seen in the shoulder
observed in the mean flipping time PDF of the solidary spins in
Fig. \ref{figure4}. For low temperatures we observed that the
shoulder does not seem to depend on temperature. Instead, the peak
of the slow spins moves to higher values in accordance with an
activation process with a characteristic energy barrier (see inset
in Fig. \ref{figure4}). This energy barrier, 4J, corresponds to
flipping a spin with only one frustrated bond. This should be
contrasted with the fact that the peak of the fast spins does not
move with temperature as shown in the inset. However, we must
point out that a possible difference could be present in the tails
of the distributions. For both, fast and slow spins, the tails
seem to be power-law like, $p_f(\tau_f)\sim\tau_f^{-1.7}$.

Summarizing, we have presented a numerical study of the
two-dimensional $\pm J$ EA spin glass focusing on how the
information of the topology of the ground state manifests in the
finite low temperature dynamics. We have concentrated in the
pre-asymptotic aging regime of this particular model as representative of
glassy dynamics.

In the ground state, spins can be divided in two sets, solidary and non-solidary spins. In the EA
model, this characterization is non-trivial and deserves careful and time consuming
simulations \cite{Ramirez}.
Once these two sets were identified, we analyzed the contribution
of each set to the finite temperature dynamics. The autocorrelation function
for each set of spin behaves differently, showing a faster initial decay for
non-solidary spins. This naturally leads to the analysis of dynamical
heterogeneities. First, we analyze the mean persistence time
distribution, and show that it is intimately related to the
two-step relaxation of the autocorrelation function. A fraction
of non-solidary spins, with lower mean persistence times,
give rise to the first decay of the autocorrelation function.
The decay observed at longer times, corresponds to a peak in
the mean persistence time distribution, and is shared by solidary and
non-solidary spins.

Finally, we test the relevance of the separation in solidary and
non-solidary spins in the mean flipping time distribution, which
presents strong dynamical heterogeneities. As was already pointed
out in Ref. \cite{Ricci} this distribution presents two sharp
peaks. This time-scale separation was used to dynamically define
groups of slow and fast spins \cite{Ricci}. Here we show for the
first time that these dynamical characterization is well accounted
for by the ground state characterization in solidary and
non-solidary spins.  Furthermore, new interesting and promising
questions arise. For instance, the mean flipping time distribution
for solidary spins presented in Fig. \ref{figure4} presents a
clear shoulder at low mean flipping times. This shoulder
corresponds to an internal structure within the set of solidary
spins. This suggests that a further division into sub-sets could
refine our results, and should be of relevance for understanding
heterogeneities in EA spin
glasses with continuous coupling distributions.\\

We thank S. A. Cannas, L. F. Cugliandolo, D. Dom\'{\i}nguez,
F. Nieto and A. J. Ramirez-Pastor
for helpful discussions and suggestions.
F.R. thanks Univ. Nac. de San Luis (Argentina) under project
322000 and Millennium Scientific Iniciative (Chile) under
contract P-02-054-F for partial support.
P.M.G. acknowledges financial support from CONICET
(Argentina), ANPCyT PICT 2003 (Argentina), Fundaci\'on Antorchas
(Argentina), ICTP NET-61 (Italy).

\end{document}